\documentclass[journal,comsoc]{IEEEtran}
\pdfoutput=1
\usepackage[T1]{fontenc}% optional T1 font encoding
\usepackage{xspace,amsmath,amssymb,amsfonts,epsfig,syntonly,bbm,dsfont,cite}
\usepackage{color,url,textcomp,empheq,boxedminipage}
\usepackage{algorithmicx,algorithm,algpseudocode}
\usepackage{empheq}
\usepackage{graphicx,graphics,subfigure}
\usepackage{multirow,multicol}
\usepackage{psfrag}
\usepackage{stfloats}
\usepackage{flushend}

\long\def\symbolfootnote[#1]#2{\begingroup
\def\thefootnote{\fnsymbol{footnote}}
\footnote[#1]{#2}\endgroup}

\psfull

\begin{document}

\title{Joint Mode Selection and Beamforming Designs for Hybrid-RIS Assisted ISAC Systems}

\author{Yingbin Lin,~\IEEEmembership{Student Member, IEEE}, Feng Wang,~\IEEEmembership{Member, IEEE}, Xiao Zhang,~\IEEEmembership{Member, IEEE},\\ Guojun Han,~\IEEEmembership{Senior Member, IEEE}
and Vincent K. N. Lau,~\IEEEmembership{Fellow, IEEE}

\thanks{Y. Lin, F. Wang, and G. Han are with the School of Information Engineering, Guangdong University of Technology, Guangzhou 510006, China (e-mail: yingbinlin0901@163.com, fengwang13@gdut.edu.cn, gjhan@gdut.edu.cn).}

\thanks{X. Zhang is with College of Computer Science, South-Central Minzu University, Wuhan 430079, China (e-mail: xiao.zhang@
my.cityu.edu.hk).}

\thanks{V. K. N. Lau is with the Department of Electronic and Computer Engineering, The Hong Kong University of Science and Technology, Hong Kong (e-mail: eeknlau@ust.hk).}

\vspace{-0.9cm}
}

\maketitle
\begin{abstract}
 This letter considers a hybrid reconfigurable intelligent surface (RIS) assisted integrated sensing and communication (ISAC) system, where each RIS element can flexibly switch between the active and passive modes. Subject to the signal-to-interference-plus-noise ratio (SINR) constraint for each communication user (CU) and the transmit power constraints for both the base station (BS) and the active RIS elements, with the objective of maximizing the minimum beampattern gain among multiple targets, we jointly optimize the BS transmit beamforming for ISAC and the mode selection of each RIS reflecting element, as well as the RIS reflection coefficient matrix. Such formulated joint hybrid-RIS assisted ISAC design problem is a mixed-integer nonlinear program, which is decomposed into two low-dimensional subproblems being solved in an alternating manner. Specifically, by using the semidefinite relaxation (SDR) technique along with the rank-one beamforming construction process, we efficiently obtain the optimal ISAC transmit beamforming design at the BS. Via the SDR and successive convex approximation (SCA) techniques, we jointly determine the active/passive mode selection and reflection coefficient for each RIS element. Numerical results demonstrate that the proposed design solution is significantly superior to the existing baseline solutions.
\end{abstract}

\vspace{-0.5cm}

\vspace{-0.5cm}
\section{Introduction}
The emerging smart internet of things (IoT) applications (such as autonomous driving and virtual reality) highly rely on the capability of integrated sensing and communication (ISAC) over wireless networks \cite{Liu_JASC22}. To enhance the ISAC performance in communication rate/coverage and sensing range/resolution, reconfigurable intelligent surface (RIS) with a large number of reflecting elements has been extensively investigated, by optimizing wireless propagation environments via adjusting the amplitudes and phases of RIS reflecting element phases with low power consumption\cite{Song_TSP23,Liao_TCOM23,Wang_TVT21,Wu_TWC19}.

%Nonetheless, pure passive RIS systems encounter specific challenges in practical scenarios, especially in long-distance and non-line-of-sight settings, leading to notable path loss attenuation and limited enhancements in performance \cite{r8, r9}. The emergence of active RIS equipped with power amplifiers holds the potential to effectively mitigate the mentioned challenges\cite{r11,r12,r13}. The authors proposed the design of an active RIS circuit and verifies the signal model, and jointly optimize the beamforming and the reflection matrix of the RIS to enhance the security of wireless transmission\cite{r11,r12}. However, these benefits come with increased power consumption and costs, along with the introduction of additional noise by the active amplifier components, further constraining design flexibility and diminishing potential applications.

Different from the fully-passive or the fully-active RIS, the hybrid active-passive RIS, comprising both passive and active reflecting elements, has been recently proposed to balance the ISAC performance enhancement and RIS hardware cost \cite{Nguyen_TVT22,Sankar_EUSIPCO22}. Typically, the hybrid active-passive RIS is equipped with switches, so that each RIS reflecting element can flexibly switch between the active and passive modes adapting to various ISAC design requirements. 
Taking into account the estimation error of the targets position, the work \cite{Liao_CL23} investigated the hybrid-RIS assisted ISAC system so as to maximize the sensing beampattern gain for multiple targets. From a rate-maximization perspective, the majorization-minimization technique \cite{Ju_TVC24} and fractional program technique \cite{Hong_CNC24} were employed for optimizing the transmit precoding matrices and the active-passive RIS reflection matrix. Aiming to maximize the secrecy rate, the work \cite{Ma_WCL24} proposed an alternate optimization based design of optimizing the transmit beamformer and the hybrid-RIS coefficients. 

%The above literature have investigated the potential of hybrid active-passive RIS in improving system performance, however, these studies typically randomize the positioning of active elements in hybrid-RIS. 
% And the work \cite{DV_CL24} extra considered the number of active elements to maximize the achievable rate of a hybrid RIS-enhanced single-input single output communication. 

%The further study \cite{Peng_CL23} determined the optimal number of active/passive elements for maximizing energy efficiency.
%Nevertheless, the optimization of each reflecting element's mode to enhance system performance remains a challenge. This is due to the discrete binary nature of the mode indicator variables, which presents a formidable obstacle to optimization and results in a loss of performance.

Note that in the existing hybrid-RIS assisted ISAC works\cite{Nguyen_TVT22,Sankar_EUSIPCO22,Wu_TWC19,Liao_CL23,Ju_TVC24,Ma_WCL24,Hong_CNC24}, the number of active and passive reflecting elements is prefixed. For energy efficiency maximization in a hybrid-RIS assisted communication system, the work \cite{Peng_CL23,Huang_TWC24} investigated the active and passive reflecting mode selection problem, which showed at most one active element is required under the Rayleigh fading channel \cite{Peng_CL23} and the Dinkelbach method is employed to optimize RIS element mode selection and beamformers in a tractable form\cite{Huang_TWC24}. Nonetheless, under the requirements of quality of services (QoS) for ISAC, it lacks the investigation of jointly determining the optimal number of active and passive reflecting elements and optimizing the BS transmit beamforming and RIS reflection matrix. 

Motivated by this, in this letter we investigate the joint passive/active mode selection and beamforming design problem for hybrid-RIS assisted ISAC system. Considering a hybrid-RIS assisted system, dual-functional base station (BS) is required to simultaneously sense multiple targets and transmit information to multiple communication users (CUs). Subject to the SINR constraints of CUs, the transmit power constraints of BS and RIS active elements, and the hybrid RIS noise power threshold constraints, our objective is to maximize the minimum sensing beampattern gain among all the targets. We jointly optimize the BS transmit beamforming vectors, the RIS mode selection, as well as the RIS reflection matrix. The joint design problem is formulated as a mixed-integer nonlinear program. As a low-complexity solution, we decompose the formulated problem into two low-dimensional subproblems, which are solved in an alternating manner. In particular, we efficiently obtain the optimal ISAC transmit beamforming vectors at the BS via the semidefinite relaxation (SDR) technique along with the rank-one beamforming construction process. By leveraging the SDR and successive convex approximation (SCA) techniques, we jointly optimize the active/passive mode selection and refection coefficient for each RIS element. The proposed algorithm is guaranteed to converge to a suboptimal solution. Numerical results demonstrate that our proposed ISAC design solution outperforms the existing baseline schemes.

\section{System Model and Problem Formulation}
\subsection{System Model}
\begin{figure}[h]
    \centering        
    \includegraphics[width=0.75\linewidth]{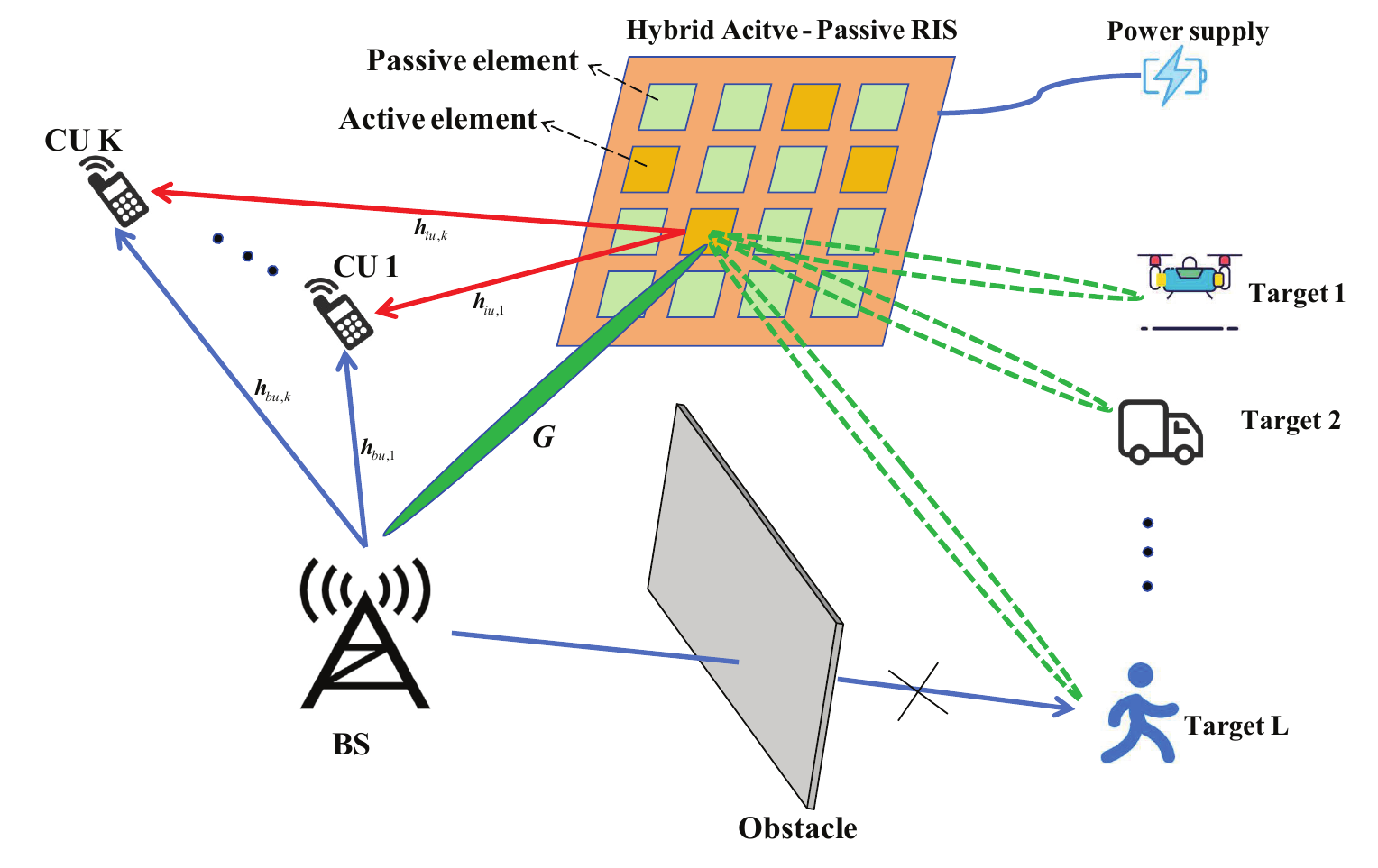}
    \vspace{-0.3cm}
    \caption{The hybrid active-passive RIS assisted ISAC system model, consisting of a number of $L$ targets and a number of $K$ CUs.}
    \label{fig1}
    \vspace{-0.3cm}
\end{figure}

As shown in Fig.~\ref{fig1}, we consider a hybrid active-passive RIS assisted ISAC system with $L$ targets and $K$ single-antenna CUs, where the BS is equipped with $M > 1$ antennas. Assisted by the hybrid-RIS, the BS is responsible for sending communication signals for $K$ CUs and dedicated radar signals for $L$ targets. The hybrid-RIS is equipped with a uniform planar array (UPA) of $N=N_{x}\times N_{y}$ elements, where $N_{x}$ and $N_{y}$ denote the number of elements of the RIS along $x$-axis and $y$-axis, respectively.

Denote by $s_{k}\in \mathbb{C}$ and $\boldsymbol{w}_{k} \in \mathbb{C} ^ {M\times 1} $ the information bearing symbol and the transmit beamforming vector for each CU $k\in \mathcal{K} \triangleq{\{1,\dots ,K}\}$, respectively. Denote by $\boldsymbol{x}_{0}\in \mathbb{C}^{M\times1}$ the zero-mean dedicated sensing signal of the BS. Hence, the ISAC transmission signal model is given as
\begin{equation}
\boldsymbol{x}=\boldsymbol{x}_{0}+\sum_{k=1}^{K}\boldsymbol{w}_{k}s_{k},
\end{equation}
where $\{s_{k}\}_{k\in \mathcal{K}}$ are assumed to be zero-mean independent and identically distributed (i.i.d.) random variables with unit variance i.e., $\mathbb{E}[|s_{k}| ^2]=1,\forall k\in\mathcal{K}$. Based on (1), the BS's transmit power is expressed as
\begin{equation}
\mathbb{E}[\|\boldsymbol{x}\| ^2]=\text{tr}(\boldsymbol{R}_{0})+\sum_{k=1}^{K}\|\boldsymbol{w}_{k}\|^2,
\end{equation}
where $\boldsymbol{R}_{0}=\mathbb{E}[\boldsymbol{x}_{0}\boldsymbol{x}_{0}^H]$ denotes the BS sensing covariance matrix.

Let $\boldsymbol{h}_{bu}\in \mathbb{C}^{M\times 1} $ and $\boldsymbol{h}_{iu}\in\mathbb{C}^{N\times 1}$ denote the fading channel vectors from BS to CU $k\in \mathcal{K}  $ and the hybrid-RIS to CU $k\in \mathcal{K} $, respectively. And $\boldsymbol{G}\in \mathbb{C}^{M\times N}$ the complex x-valued channel matrix from the BS to the RIS. Note that via the existing pilot-aid training based or uplink-downlink channel reciprocity based channel estimation methods, the channel state information (CSI) can be obtained for the RIS-assisted ISAC systems \cite{r19}. In this letter, we assume CSI is perfectly obtained by the BS and neglect the CSI estimation error for facilitating the system designs.

%Denote by $N_{a}$ and $(N-N_a)$ the number of active and passive elements in the hybrid-RIS, where $0\leq N_a\leq N$ needs to optimized. Different from the passive RIS elements whose amplitudes are fixed, both the amplitude of phase of each active RIS element can be simultaneously optimized. We now introduce notations to mathematically model the hybrid-RIS. 

Denote by $\boldsymbol{Q} = \text{diag}(q_1, ..., q_N )$ the RIS's passive-active mode selection matrix, where $q_n\in\{0,1\}$ is the model selection variable of each RIS element $n\in{\cal N}\triangleq\{1,...,N\}$. Specifically, if $q_n=0$, then RIS element $n$ stays in the passive reflection mode, and if $q_n=1$, then RIS element $n$ stays in the active reflection mode. Denote by $\boldsymbol{\Phi}= \text{diag}(\beta_{1}e^{j\theta_1},...,\beta_{N}e^{j\theta_N})$ the RIS reflection coefficient matrix, where $\beta_{n}e^{j\theta_{n}}$ is termed as the reflection coefficient of each RIS element $n\in{\cal N}$ with $\beta_n$ and $\theta_n$ respectively denoting its amplitude and phase shift. For the $n$-th RIS element with $q_n=1$, its amplitude $\beta_n$ and phase $\theta_n$ can be simultaneously optimized. By contrast, for the $n$-th RIS element with $q_n=0$, its amplitude is fixed as $\beta_n=1$ and only the phase $\theta_n$ can be optimized. 
As such, we have
\begin{equation}
\left\{
\begin{array}{ll}
\beta_n\in [0,\beta_{\max}]~~\text{if}~  q_n=1 \\   \beta_n=1 ~~~~~~~~~~~\text{if}~q_n=0,
\end{array}
\right. 
\end{equation}
where $\beta_{\max}>1$ denotes the maximum amplifying threshold for an RIS element in the active reflection mode. Equivalently, for the mode selection of each RIS element $n\in{\cal N}$, we have the constraint as $1-q_n\le \beta_{n}\le 1-q_n+q_n\beta_{\max}$.

Note that, when an RIS element $n\in{\cal N}$ stays in the passive mode, the transmission power is not required for this element. By contrast, when an RIS element $n\in{\cal N}$ stays in the active mode, this element requires an additional transmission power for adapting to its amplitude value $\beta_n$ \cite{Liao_CL23}. Therefore, the hybrid RIS's total transmission power is expressed as
\begin{subequations}
\begin{align}
    P^{\text{ris}} &=\mathbb{E}[\|\boldsymbol{\Phi} \boldsymbol{Q}\boldsymbol{G}\boldsymbol{x}+\boldsymbol{\Phi}\boldsymbol{n}_{\text{ris}})\|^2] \\
    &= 
\|\boldsymbol{\Phi}\boldsymbol{Q}\boldsymbol{G}\boldsymbol{x}\|^2+\|\boldsymbol{\Phi} \boldsymbol{Q})\|_{F}^2\sigma_{\text{ris}}^2,
 \end{align}
 \end{subequations}
where $\boldsymbol{n}_{\text{ris}} \sim
\mathcal{CN} (0,\sigma_{\text{ris}}^2\boldsymbol{Q})$ is the additive white Gaussian noise vector incurred by the RIS's elements in the active mode, and $\|\boldsymbol{A}\|_F$ denotes the Frobenius norm of matrix $\boldsymbol{A}$.

In the hybrid-RIS assisted ISAC system under consideration, each CU $k\in\mathcal{K} $ receives signals from the BS via both the line-of-sight link and the BS-RIS-CU link. As a result, the received signal of CU $k\in\mathcal{K} $ is modeled as 
\begin{align}
y_{k}=\boldsymbol{h}_{\text{CU},k}^H\boldsymbol{x}+\boldsymbol{h}_{iu,k}^H\boldsymbol{\Phi}\boldsymbol{Q}\boldsymbol{n}_{\text{ris}}+n_{k}, 
\end{align}
where $\boldsymbol{h}_{\text{CU},k}\triangleq\boldsymbol{G}^H\boldsymbol{\Phi}h_{iu,k}+h_{bu,k}$ denotes the equivalent channel vector from the BS to CU $k$, and $n_{k}\sim\mathcal{CN} (0,\sigma^2_{k})$ denotes the additive white Gaussian noise at CU $k$'s receiver. Note that the BS's sensing signal $\boldsymbol{x}_{0}$ is deterministic and thus available to the $K$ CUs. As such, by removing the interference caused by the BS's sensing signal $\boldsymbol{x}_{0}$, the SINR of each CU $k\in\mathcal{K}$ is given as 
\begin{equation}
{\rm{SINR}}_k=\frac{|\boldsymbol{h}_{\text{CU},k}^H\boldsymbol{w}_{k}|^{2}} 
    {\textstyle \sum_{j\ne k}^{K}|\boldsymbol{h}_{\text{CU},k}^H\boldsymbol{w}_{j}|^2    +\|\boldsymbol{h}_{iu}^H\boldsymbol{Q}\boldsymbol{\Phi}\|^2\sigma_{\text{ris}}^2+\sigma^2_{k}}.
\end{equation}

%\subsection{Sensing Beampattern Gains for the L Targets }
% Let $\theta_l\in(-\frac{2}{\pi},\frac{2}{\pi})$ and $\varphi_l\in(-\frac{2}{\pi},\frac{2}{\pi}) $ denote the azimuth
% angle and the elevation angle of target $l\in \mathcal{L}$ with respect
% to the RIS, respectively. 

Denote by $\theta_l\in (-\frac{2}{\pi}, \frac{2}{\pi})$ and $ \varphi_l\in (-\frac{2}{\pi},\frac{2}{\pi})$ the azimuth angle and elevation angle of target $l\in\mathcal{L}$ with respect to the RIS, respectively. Denote by $d_x$ and $d_y$ the distances between two consecutive RIS elements along the $x$-axis and $y$-axis, respectively. For each target $l\in \mathcal{L}$ with the azimuth and elevation-angle pair $(\theta_l,\varphi_l)$, the RIS's array response matrix is denoted as $\boldsymbol{\alpha}(\theta_l,\varphi_l)$, which is expressed as
  \begin{align}
    &\boldsymbol{\alpha}(\theta_l,\varphi_l)=\left[1,e^{j\frac{2\pi d_{x}}{\lambda}\sin\theta_l \cos\varphi_l},...,e^{j\frac{2\pi(N_{x}-1) d_{x}}{\lambda}\sin\theta_l \cos\varphi_l}\right]^T  \notag \\
    &\otimes \left[1,e^{j\frac{2\pi d_{y}}{\lambda}\sin\theta_l \sin\varphi_l},...,e^{j\frac{2\pi(N_{y}-1) d_{y}}{\lambda}\sin\theta_l \sin\varphi_l}\right]^T,
 \end{align}
where $\lambda$ represents the wavelength of the radio-frequency transmit signal of the BS, and the operator $\otimes$ denotes the Kronecker product. Since that both the communication signals $\{s_{k}\}_{k\in \mathcal{K}}$ and the sensing signal $\boldsymbol{x}_0$ can concurrently illuminate the $L$ targets for sensing purposes, the sensing beampattern gain $\rho\left(\theta_{l}, \varphi_{l}\right)$ for each target $l\in \mathcal{L}$ is expressed as
\begin{equation}
    \rho\left(\theta_{l}, \varphi_{l}\right)=\boldsymbol{h}_{l}^{H}\big(\boldsymbol{R}_{0}+\sum_{k=1}^{K} \boldsymbol{w}_{k} \boldsymbol{w}_{k}^{H}\big) \boldsymbol{h}_{l},
\end{equation}
where $\boldsymbol{h}_{l} \triangleq \boldsymbol{G}^{H} \boldsymbol{\Phi}^{H} \boldsymbol{\alpha}\left(\theta_{l}, \varphi_{l}\right)$ represents the cascaded BS-RIS-target channel vector from the BS to each target $l\in{\cal L}$ via the RIS. Due to the noise $\boldsymbol{n}_{\text{ris}}$ of the RIS elements in the active mode, the additional noise power generated at each target $l\in \mathcal{L}$ is denoted as $\xi_{l}^{\text {ris }}$, which is expressed as
\begin{subequations}
\begin{align}
\xi_{l}^{\text {ris }} &=\mathbb{E}\left[\| \boldsymbol{\alpha}^{H}\left(\theta_{l}, \varphi_{l}\right)\boldsymbol{\Phi}  \boldsymbol{n}_{\text {ris }} \|^{2}\right]  \\
&=   \| \boldsymbol{\alpha}^{H}\left(\theta_{l}, \varphi_{l}\right)\boldsymbol{Q} \boldsymbol{\Phi}  \|^{2} \sigma_{\text {ris }}^{2}.
\end{align}
\end{subequations}

\subsection{Problem Formulation }
Aiming to maximize the minimum sensing beampattern gain among the $L$ targets, we jointly optimize the BS transmit beamforming vectors $\{\boldsymbol{w}_{k}\}_{k\in \mathcal{K}}$ for communication and covariance matrix $\boldsymbol{R}_0$ for sensing, as well as the RIS mode selection variables $\{q_n\}_{n\in{\cal N}}$ and reflection coefficient matrix $\boldsymbol{\Phi} $. As such, the joint RIS mode selection and ISAC beamforming design problem under consideration is formulated as 
\begin{subequations}\label{p1}
\begin{align}
        ({\rm P1}):  &  \max _{\left\{\boldsymbol{w}_{k}\right\}_{k=1}^{K}, \boldsymbol{R}_{0}\succeq\boldsymbol{0}, \boldsymbol{\Phi},\{q_n\}_{n\in{\cal N}}} \min _{l \in \mathcal{L}} ~~\rho(\theta_{l}, \varphi_{l})\\
        \text { s.t. }& \sum_{k=1}^{K}\left\|\boldsymbol{w}_{k}\right\|^{2}+\operatorname{tr}\left(\boldsymbol{R}_{0}\right) \leq P_{0}\\
       % & \boldsymbol{R}_{0} \succeq \mathbf{0}\\
        &\mathrm{SINR}_{k} \geq \Gamma_{k}, \forall k \in \mathcal{K} \\
        & 0\leq P^{\text {ris }} \leq P_{\max }^{\text {ris }}\\
        &\xi_{l}^{\text {ris }} \leq \xi_{\max }^{\text {ris }}, \forall l\in{\cal L}\\
        & 0<\theta_{n} \leq 2 \pi,\forall n \in \mathcal{N} \\ 
        &1-q_n\le \beta_{n}\le 1-q_n+q_n\beta_{\max }\\
        &q_n\in\{0,1\},\forall n\in \mathcal{N},
\end{align}
\end{subequations}
where (\ref{p1}b) denotes the BS's transmit power constraint with $P_{0}$ being the BS transmit power budget; (\ref{p1}c) denotes the SINR constraint for CU $k$ with $\Gamma_{k}$ being its minimal SINR threshold; (\ref{p1}d) denotes the RIS transmit power constraint with $P_{\max }^{\text {ris}}$ being its total power budget; (\ref{p1}e) denotes the RIS noise power constraint for each target $l\in{\cal L}$ with $\xi_{\max }^{\text {ris}}$ denoting its maximal power threshold; (\ref{p1}f) and (\ref{p1}g) denote the phase and amplitude constraints of each RIS element $n\in{\cal N}$, respectively; (\ref{p1}h) denotes the active/passive mode selection of RIS element $n$. Due to the discreteness of the RIS mode selection variables $\{q_n\}_{n\in{\cal N}}$ and the coupling of design variables, (P1) is a mixed-integer nonlinear program (MINLP).

\vspace{-0.4cm}
 \section{Proposed Design Solution}
 In this section, we propose an alternating optimization based solution for (P1), where the BS beamforming design variables $(\left\{\boldsymbol{w}_{k}\right\}_{k=1}^{K},\boldsymbol{R}_{0})$ and the RIS design variables $(\{q_n\}_{n\in{\cal N}},\boldsymbol{\Phi})$ are iteratively optimized in an alternating manner.

 \subsection{Optimization of $(\left\{\boldsymbol{w}_{k}\right\}_{k=1}^{K},\boldsymbol{R}_{0})$ Under Given $(\{q_n\}_{n\in{\cal N}},\boldsymbol{\Phi})$}

First, define $\boldsymbol{W}_k\triangleq\boldsymbol{w}_k\boldsymbol{w}_k^H\succeq 0 $ for each CU $\forall k \in \mathcal{K} $, and it follows that $\text{rank}(\boldsymbol{W}_k)\leq 1$, $\forall k \in \mathcal{K}$. By substituting $\{\boldsymbol{W}_k\}_{k\in{\cal K}}$ in (P1), the problem of optimizing $(\{\boldsymbol{W}_{k}\}_{k=1}^K,\boldsymbol{R}_{0})$ under the given RIS mode selection variables and reflection coefficient matrix $(\{q_n\}_{n\in{\cal N}},\boldsymbol{\Phi})$ is formulated as
\begin{subequations}
\begin{align}
&  ({\rm P1.1}): \max _{\left\{\boldsymbol{W}_{k} \right\}_{k=1}^{K}, \boldsymbol{R}_{0}}~~ \rho'\\
 & \text { s.t. } \text{tr}\left(\boldsymbol{R}\boldsymbol{h}_l\boldsymbol{h}_l^H\right) \geq \rho',~\forall l\in {\cal L}\\
  &\text{tr}(\boldsymbol{R})\leq P_0\\
  &\text{tr}\left(\left(\frac{1+\Gamma_{k}}{\Gamma_{k}}\boldsymbol{W}_k\boldsymbol{h}_{\text{CU},k}\boldsymbol{h}_{\text{CU},k}^H\right)-\boldsymbol{\bar{W}}_{k}\right) \geq b_k ,\forall k\in{\cal K}\\
  & \text{tr}\left(\boldsymbol{\Phi}\boldsymbol{Q}\boldsymbol{G}\boldsymbol{R}\boldsymbol{G}^H\boldsymbol{Q}^H\boldsymbol{\Phi}^H\right)+\|\boldsymbol{\Phi} \boldsymbol{Q})\|_{F}^2\sigma_{\text{ris}}^2\leq P_{\max }^{\text {ris}}\\
 & \boldsymbol{R}_{0} \succeq \mathbf{0},\boldsymbol{W}_k\succeq 0 ,\text{rank}(\boldsymbol{W}_k)\leq 1, \forall k\in{\cal K}, 
\end{align}
\end{subequations}
where $\boldsymbol{R}\triangleq\boldsymbol{R}_{0}+\sum_{i=1}^{K}\boldsymbol{W}_i$, $\boldsymbol{\bar
W}_k \triangleq \sum_{i=1}^{K}\boldsymbol{W}_i\boldsymbol{h}_{\text{CU},k}\boldsymbol{h}_{\text{CU},k}^{H}$, and $b_k\triangleq\|\boldsymbol{h}_{iu}^H\boldsymbol{\Phi}\boldsymbol{Q}\|^2\sigma_{\text{ris}}^2+\sigma^2_{k}$, $\forall k\in{\cal K}$.

Note that due to the rank-one constraints of $\{\boldsymbol{W}_k\}_{k\in{\cal K}}$, problem (P1.1) is still non-convex. Nonetheless, such rank-one constraints can be relaxed by employing the well-known semidefinite relaxation (SDR) technique \cite{2004Convex}, where (P1.1) is relaxed as a convex semidefinite programming (SDP) problem. Denote by $(\{\boldsymbol{W}_k^{*}\}_{k\in{\cal K}}, \boldsymbol{R}_0^{*})$ the optimal solution to the rank-relaxed problem (P1.1). If $\text{rank}(\boldsymbol{W}_k^{*})\leq 1$, $\forall k\in{\cal K}$, then $(\{\boldsymbol{W}_k^{*}\}_{k\in{\cal K}}, \boldsymbol{R}_0^{*})$ is optimal for the original problem (P1.1). On the other hand, if $\text{rank}(\boldsymbol{W}_k^{*}) > 1$ for a certain CU $k\in{\cal K}$, an optimal solution $(\{\tilde{\boldsymbol{w}}_k\}_{k=1}^K,\tilde{\boldsymbol{R}}_0)$ is constructed for (P1.1) \cite{Liao_CL23}, where $\tilde{\boldsymbol{w}}_k= \frac{1}{\sqrt{\boldsymbol{h}^H_{\text{CU},k}\boldsymbol{W}_k^{*}\boldsymbol{h}_{\text{CU},k}}} \boldsymbol{W}_k^{*}\boldsymbol{h}_{\text{CU},k}$, $\forall k\in{\cal K}$, and $\tilde{\boldsymbol{R}}_0=\boldsymbol{R}_0^{*}+\sum_{k=1}^{K}\boldsymbol{W}_k^{*}- \sum_{k=1}^{K}\tilde{\boldsymbol{w}}_k\tilde{\boldsymbol{w}}_k^H$.

\subsection{Optimization of $(\{q_n\}_{n\in{\cal N}},\boldsymbol{\Phi})$ Under Given $(\left\{\boldsymbol{w}_{k}\right\}_{k=1}^{K},\boldsymbol{R}_{0})$}

 To start with, we define ${\boldsymbol{v}}\triangleq [\Phi_{1,1},...,\Phi_{N,N},1]^T\in\mathbb{C}^{(N+1)\times 1}$, $\boldsymbol{V}\triangleq {\boldsymbol{v}}{\boldsymbol{v}}^H$. 
 Note that the mode selection matrix $\boldsymbol{Q}$ and the RIS reflection matrix $\boldsymbol{\Phi}$ are highly coupled together \cite{Huang_TWC24}. As such, we introduce an auxiliary matrix $\boldsymbol{Z}$. The entry $Z_{i,j}$ needs to satisfy the double-sided linear constraints as
\begin{subequations}\label{c1}
\begin{align}
&-q_{i}\tilde{c}\leq Z_{i,j}\leq q_{i}\tilde{c} \\
%&Z_{i,j}\leq Q_{i,i}\tilde{c}, \\
&-q_{i}\tilde{c}\leq Z_{j,i} \leq  q_{i}\tilde{c} \\
%%&Z_{j,i}\leq Q_{i,i}\tilde{c}, \\
&V_{i,j}-(1-q_{i})\tilde{c}\leq Z_{i,j} \leq V_{i,j}+(1-q_{i})\tilde{c}, 
%&Z_{i,j}\leq U_{i,j}+(1-Q_{i,i})\tilde{c}, \forall i,j\in \mathcal{N}
\end{align}
\end{subequations}
where $i\in{\cal N}$, $j\in{\cal N}$, and $\tilde{c} > 0$ is a constant with a sufficiently large value. It is verified that (\ref{c1}) guarantees $\boldsymbol{Z} = \boldsymbol{Q}\boldsymbol{V}(1:N,1:N)\boldsymbol{Q}^H$, i.e., $Z_{i,j}=q_{i}V_{i,j}q_{j}$, $\forall i,j\in{\cal N}$.

With $\boldsymbol{V}$ and $\boldsymbol{Z}$ at hand, the RIS amplitude constraints in (\ref{p1}g) are equivalently re-expressed as 
\begin{subequations}\label{c2}
\begin{align}
& V_{n,n}\leq \beta^2_{\max}\\
& V_{n,n}-Z_{n,n} = 1 - q_{n},
\end{align}
\end{subequations}
where $n\in{\cal N}$.

%In addition, a modulus constraint should be imposed on the passive element to limit the amplitude:
%\begin{align}
 %  [(\boldsymbol{I}-\boldsymbol{Q})\boldsymbol{U}]_{nn}=[\boldsymbol{I}-\boldsymbol{Q}]_{nn}
%\end{align}
%But here, for the algorithm to hold, this restriction is relaxed using a similar approach as in the previous study
 % \begin{align}\label{c3}
  % [\boldsymbol{U}-\boldsymbol{Z}]_{nn}\leq[\boldsymbol{I}-\boldsymbol{Q}]_{nn}
%\end{align}

%Further, the binary constraint (10i) can be rewritten:
%%   q_n(1-q_n)=0,\forall n \in \mathcal{N}
%\end{align}
%It is worth noting that the optimization variable $q_n$ for pattern scheduling is binary discrete, so it involves integer constraints, which makes the RIS configuration bitter. 

Next, for the binary RIS active-passive mode selection constraint (\ref{p1}h), we employ the following relaxation as
\begin{align}\label{c4}
    0\le q_n\le 1,\forall n \in \mathcal{N}.
\end{align}
Furthermore, in order to reduce the gap due to the relaxation of $\{q_n\}_{n\in{\cal N}}$, the function $\Lambda=\sum_{n=1}^N q_{n}(1-q_{n})$ is used as a penalty term. Nonetheless, $\Lambda$ is a non-convex function with respect to $\{q_n\}_{n\in{\cal N}}$. As such, via the SCA technique, we introduce a upper-bound of $\Lambda$ as a convex penalty term $\tilde{\Lambda}$, i.e., $\tilde{\Lambda} = \sum_{n=1}^N ( q_{n}-(q_{n}^{(i)})^{2}-2q_{n}^{(i)}(q_{n}-q_{n}^{(i)}))$, where $\{q_{n}^{(i)}\}_{n\in{\cal N}}$ are the obtained RIS mode selection variables at the $i$-th iteration. For notational convenience, we define the following symbols as
\begin{subequations}
\begin{align} \bar{\boldsymbol{R}}^{\text{tar}}_l&=[\boldsymbol{R}^{\text{tar}}(\theta_l,\varphi_l),\boldsymbol{0};\boldsymbol{0},0],~\forall l\in{\cal L} \\ 
\bar{\boldsymbol{R}}_{k,j}^{\text{CU}}&=[\boldsymbol{R}_{k,j}^{\text{CU}},\boldsymbol{b}_{k,j}^{\text{CU}};(\boldsymbol{b}_{k,j}^{\text{CU}})^H,0],~\forall k,j\in{\cal K}\\ 
\boldsymbol{R}_l^{\text{tar}}&\triangleq \text{diag}(\boldsymbol{\alpha}(\theta_l,\varphi_l))^H\boldsymbol{G}\boldsymbol{R}\boldsymbol{G}^H\text{diag}(\boldsymbol{\alpha}(\theta_l,\varphi_l))\\
\boldsymbol{R}_{k,j}^{\text{CU}}&\triangleq \text{diag}(\boldsymbol{h}_{iu,k})^H\boldsymbol{G}\boldsymbol{W}_j\boldsymbol{G}^H\text{diag}(\boldsymbol{h}_{iu,k})\\
\boldsymbol{b}^{\text{CU}}_{k,j} &\triangleq \text{diag}(\boldsymbol{h}_{iu,k})^H\boldsymbol{G}\boldsymbol{W}_j\boldsymbol{h}_{bu,k}\\
\boldsymbol{P}^{\text{ris}} &\triangleq  \boldsymbol{G}\boldsymbol{R}\boldsymbol{G}^H + \sigma_{\text{ris}}^2\boldsymbol{I}_N\\
\boldsymbol{\Sigma}^{\text{ris}}_{k} &\triangleq \sigma_{\text{ris}}^2 \text{diag}(\boldsymbol{h}_{iu,k})^H \text{diag}(\boldsymbol{h}_{iu,k}),\forall k\in{\cal K}\\
\boldsymbol{P}_l^{\text{tar} } &\triangleq \sigma_{\text{ris}}^2 \text{diag}(\boldsymbol{\alpha}(\theta_l,\varphi_l))^H\text{diag}(\boldsymbol{\alpha}(\theta_l,\varphi_l)),\forall l\in{\cal L}.
\end{align}
\end{subequations}
 
 As a result, with $(\rho'',\boldsymbol{V},\boldsymbol{Z},\{q_n\}_{n\in{\cal N}})$ as the design variables under consideration, the joint optimization of the RIS reflection coefficient matrix $\boldsymbol{\Phi}$ and model selection variables $\{q_n\}_{n\in{\cal N}}$ under the given $(\{\boldsymbol{w}_{k}\}_{k=1}^{K},\boldsymbol{R}_{0})$ is formulated as a convex SDP problem as
 \begin{subequations}
 \begin{align}
   & (\text{P1.2}):~\max_{\rho'',\boldsymbol{V},\boldsymbol{Z},\{q_n\}_{n\in{\cal N}}}~ ~\rho''-\mu \tilde{\Lambda}\\
   &\text{s.t.}~ \text{tr}(\bar{\boldsymbol{R}}^{\text{tar}}_l\boldsymbol{V})\geq \rho'',~l\in \cal{L}\\
    &~ \text{tr}\Big((\frac{\bar{\boldsymbol{R}}^{\text{CU}}_{k,k}}{\Gamma_k}-\sum_{j=1,j\neq k}^K \bar{\boldsymbol{R}}^{\text{CU}}_{k,j} ){\boldsymbol{V}}\Big) -\text{tr}({\boldsymbol{\Sigma}}^{\text{ris}}_{k}\boldsymbol{Z})\geq c_k,\forall k\in{\cal K}\\
    &~ \text{tr}\big({\boldsymbol{P}}^{\text{ris}}\boldsymbol{Z}\big) \leq P_{\max}^{\text{ris}},~\text{tr}\big({\boldsymbol{P}}^{\text{tar}}_l\boldsymbol{Z}\big) \leq \xi_{\max}^{\text{ris}},~\forall l\in \cal{L}\\
    &(\ref{c1}),(\ref{c2}),(\ref{c4}),
 \end{align}
 \end{subequations}
where $c_k\triangleq \sum_{j=1,j\neq k}^K |\boldsymbol{h}^H_{bu,k}\boldsymbol{w}_j|^2 -\frac{1}{\Gamma_k}|\boldsymbol{h}_{bu,k}^H\boldsymbol{w}_k|^2 + \sigma_k^2$, $\forall k\in{\cal K}$, and $\mu>0$ denotes the penalty factor. Due to the convexity, (P1.2) can then be solved efficiently by convex solvers. Denote by $(\rho''^*,\boldsymbol{V}^*,\boldsymbol{Z}^*,\{q^*_n\}_{n\in{\cal N}})$ the optimal solution to (P1.2). If $\text{rank}(\boldsymbol{V}^*)>1$, we resort to the Gaussian randomization method \cite{Wu_TWC19} to generate a suboptimal solution $(\hat{\rho}'',\hat{\boldsymbol{V}},\boldsymbol{Z}^*,\{q^*_n\}_{n\in{\cal N}})$ for (P1.2), where $\text{rank}(\hat{\boldsymbol{V}})=1$. With the solution of (P1.2), we are ready to obtain the optimized mode selection $q_n$ and coefficient $\Phi_{n,n}$ of each RIS's element $n\in{\cal N}$.

%Specifically, for each Gaussian randomization indexed by, $i=1,...,L_{\text{Gau}}$, a vector generated as $\boldsymbol{r}_i\sim \mathcal{CN}(\boldsymbol{0},\boldsymbol{V}^*)$, which is then utilized in constructing a rank-one solution. Here, the matrix $\hat{\boldsymbol{V}}= \hat {\boldsymbol{v} }\hat{\boldsymbol{v}}^H$, where $\hat{\boldsymbol{v}}_i= \text{diag}(\boldsymbol{V}^*)^{\frac{1}{2}}e^{j\arg([\frac{\boldsymbol{r}_i(1:N+1)}{{r_i}(N+1)}])}$.

%Specifics can be found in

% For each Gaussian randomization, $i=1,...,L_{\text{Gau}}$, a vector generated by $r_i\sim \mathcal{CN}(\boldsymbol{0},\boldsymbol{V}^*)$. This vector is then used to construct a rank-one solution, which is denoted by the matrix $\tilde{\boldsymbol{V}}= \tilde {\boldsymbol{v} }\tilde{\boldsymbol{v}}^H$, where $\tilde{\boldsymbol{v}}_i= \text{diag}(\boldsymbol{V}^*)^{\frac{1}{2}}e^{j\arg([\frac{\boldsymbol{r}_i(1:N+1)}{{r_i}(N+1)}])}$
%  The objective value is approximated as the maximum value among all random realisations, and thus $L_{\text{Gau}}$ is set to a sufficiently large value to guarantee that the objective value increases at each iteration. The optimal solution for (P1.2) obtained via the aforementioned Gaussian randomization process is denoted by the variable $\tilde{\boldsymbol{v}}^{\max}$. The resulting optimized RIS reflecting coefficient matrix, $\boldsymbol{\Phi} = \text{diag}(\tilde{\boldsymbol{v}}^{\max}(1:N))$, is then calculated under the given constraints of $\{\boldsymbol{w}_k\}_{k\in{\cal K}},\boldsymbol{R}_0$.

\begin{algorithm}[t]
\caption{Proposed algorithm for solving (P1)}
    \begin{algorithmic}[1]
    \State {\bf Initialize} $(\{q_n\}_{n\in{\cal N}}^{(0)},\boldsymbol{Q}^{(0)})$ and the iteration index $i = 1$;
       \State {\bf Repeat:}
            \State Obtain $(\{\boldsymbol{w}_k^{(i)}\},\boldsymbol{R}_0^{(i)})$ by (P1.1) with $(\{q_n^{(i-1)}\},\boldsymbol{\Phi}^{(i-1)})$;
            \State Obtain $(\{q_n^{(i)}\},\boldsymbol{\Phi}^{(i)})$ by (P1.2) with $(\{\boldsymbol{w}_k^{(i)}\},\boldsymbol{R}_0^{(i)})$;
            \State  Update $i = i+1$;
       \State {\bf Until:} The objective value for (P1) reaches convergence.
    \end{algorithmic}
\end{algorithm}

\vspace{-0.4cm}
\subsection{Proposed Algorithm and Analysis}
\begin{figure}
	\begin{minipage}{0.49\linewidth}
		\vspace{3pt}	
        \centerline{\includegraphics[width=\textwidth]{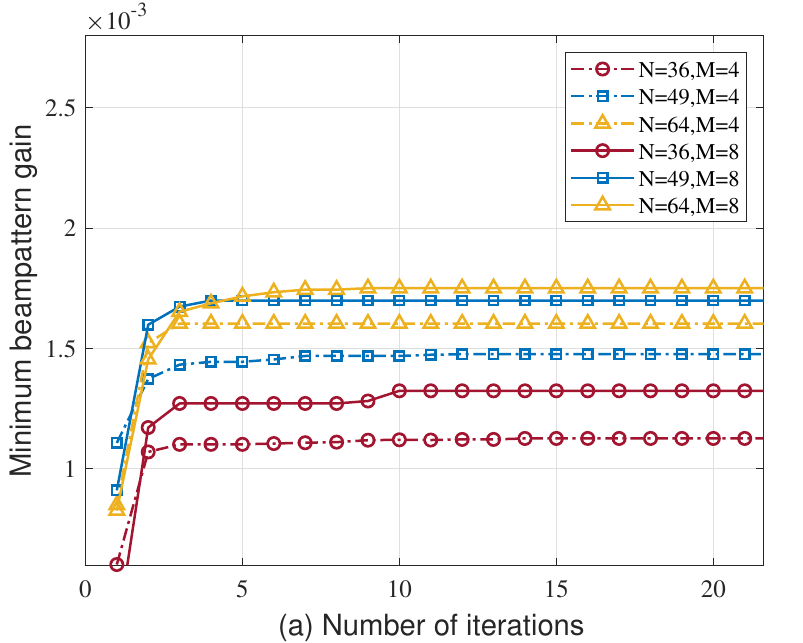}}
	\end{minipage}
	\begin{minipage}{0.5\linewidth}
		\vspace{3pt}		\centerline{\includegraphics[width=\textwidth]{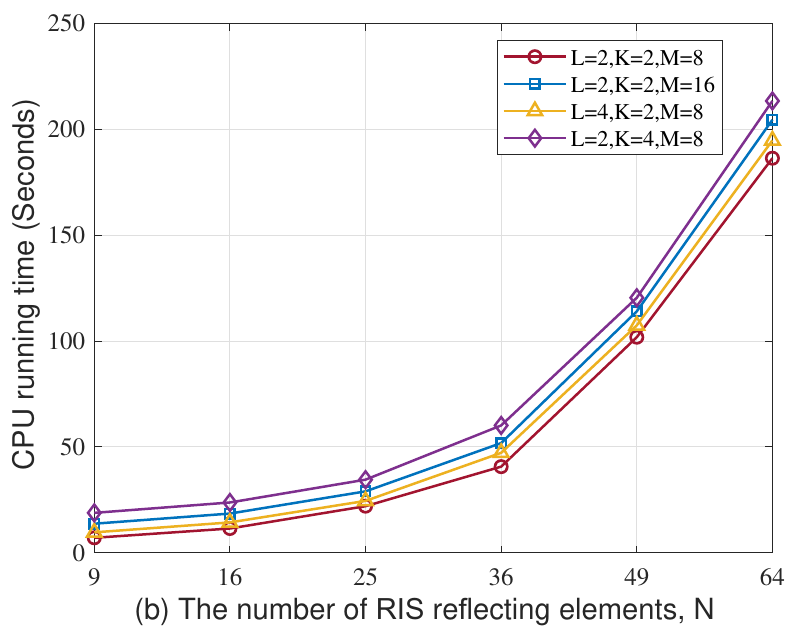}} 
	\end{minipage}
	\caption{(a) The minimum target sensing beampattern gain; 
 (b) the running time of proposed algorithm}
	\label{fig2}
    \vspace{-0.2cm}
\end{figure}

As a summary, Algorithm~1 is presented, in which problems (P1.1) and (P1.2) are alternatively solved during each iteration round. For (P1.1), its optimal beamforming solution can be efficiently obtained by the optimal rank-one solution construction process. For (P1.2), the penalty factor $\mu$ and the number of Gaussian randomizations are set to be large values so as to reduce the approximation error. It thus guarantees the objective value of (P1.2) to be non-decreasing over each iteration \cite{Liao_TCOM23}. Hence, the proposed Algorithm~1 is shown to converge to a local optimal solution of (P1). Note that per iteration, the computational complexity of solving (P1.1) is $\mathcal{O}((L+K)(K+1)^2M^4)$, where $(K+1)M^2$ and $L+K$ denote the number of optimization variables and constraints in (P1.1), respectively, and the computational complexity of solving (P1.2) is $\mathcal{O}(((2N^2+3N+2)^{2}(N^2+2(N+L)+K))+(2N^2+3N+2)^{0.5})$, where $\mathcal{O}((2N^2+3N+2)^{0.5})$ denotes the computation cost of using SCA. As a result, the computational complexity of Algorithm~1 is given as $\mathcal{O}(J(((K+1)M^2)^{2}(L+K))\log(1/\epsilon)) + \mathcal{O}(J((2N^2+3N+2)^{2}(N^2+2(N+L)+K)+(2N^2+3N+2)^{0.5})\log(1/\epsilon)) + \mathcal{O}(JL_{\text{Gau}}(N+1)^{2})$\cite{2004Convex}, where $J$ denotes the iteration number of Algorithm~1, $\epsilon$ denotes the tolerable error level, and $L_{\text{Gau}}$ denotes the number of Gaussian randomizations for (P1.2).

%The complexity of solving the problem in each internal iteration is $\mathcal{O}((N^{2}M)\log(1/\epsilon))$, where $N$ and $M$ denote the number of optimization variables and constraints \cite{2004Convex}.
%In addition, using SCA will add additional costs: $\mathcal{O}((N_1)^{1/2}\log(1/\epsilon))$, where $N_1$ denotes the number of variables.

Fig.~\ref{fig2} illustrates the convergence performance and the running time of the proposed Algorithm 1, where the system parameters are set the same as in section IV. In Fig.~\ref{fig2}(a), Algorithm 1 is observed to converge within approximately 10 iterations under different setups of the RIS element number $N$ and BS antenna number $M$. Since the number of degrees of design freedom increases with the BS antenna $M$, the minimum sensing beampattern gain increases as $M$ increases. In Fig.~\ref{fig2}(b), the running time of the proposed Algorithm 1 is observed to increase with the increasing of $N$ and $M$, as well as the target number $L$. This is expected, since a parameter-setup of $(N,M,L)$ with larger values leads to an optimization problem with higher dimensions.

\vspace{-0.4cm}
\section{Numerical Results}
In this section, we evaluate the performance of the proposed hybrid-RIS assisted ISAC design. In simulations, the BS and RIS are located at the coordinates (0, 0, 2.5) m and (20, 5, 2.5) m, respectively. The BS antenna number and the RIS element number are set as $M=8$ and $N=64$, respectively. The transmission power budget of BS and RIS is set as $P_0=0.3$ W and $P_{\max}^{\text{ris}}=-3$ dBm, respectively. We consider a Rician fading channel model with a Rician factor of 0.5 and Rayleigh fading channels for each BS-CU link. The pathloss model is $K_0(\frac{d}{d_0})^{-\alpha}$, where $K_0 = -30$ dB, $d_0 = 1$ m, and we set the pathloss exponential values of the parameter $\alpha$ as 2.5, 2.5, and 2.2 for the RIS-CU, BS-RIS, and BS-CU links, respectively. Unless specified otherwise, we consider $K=2$ CUs and $L=2$ targets, where the target azimuth and elevation-angle pair $(\theta_l,\varphi_l)$ are set as $(-60^{\circ},60^{\circ})$ and $(-30^{\circ},30^{\circ})$. The active element noise power is set to $\sigma_{\text{ris}}^2=-70$ dBm and the noise maximal power threshold is set to $\xi_{\max }^{\text {ris}}=-10$ dBm. For all the CUs, the receiver noise power and SINR threshold are set as $\sigma^2_{k}=\sigma^2=-80$ dBm and $\Gamma_k=\Gamma=5$ dB, respectively. The number of Gaussian randomization process is set to $L_{\text{Gau}} = 10^4$. For comparison, we consider the following three baseline schemes.

%Scheme 1  (random-phase scheme): In this scheme, the phase of the RIS elements is randomly determined without optimization;
\begin{itemize}
  \item Scheme 1 (Fixed-mode RIS): A number $N_a=12$ of RIS elements is fixed to be in the active mode, while the remaining RIS elements stay in the passive mode.
\item Scheme 2 (Full passive RIS): All RIS elements are in the passive mode, i.e., $q_n=0$, $\forall n\in{\cal N}$.
\item Scheme 3 (Full active RIS): All RIS elements are in the passive mode, i.e., $q_n=1$, $\forall n\in{\cal N}$.
\end{itemize}
% \begin{figure}[t]
% \centering  %图片全局居中
% \subfigure[image 1]{
% \includegraphics[width=0.33\linewidth]{picture/2.eps}}\subfigure[image 2]{
% %\label{Fig.sub.2}
% \includegraphics[width=0.33\linewidth]{picture/3.eps}}\subfigure[image 2]{
% %\label{Fig.sub.2}
% \includegraphics[width=0.33\linewidth]{picture/4.eps}}
% \caption{Insert two pictures side by side}
% \label{1}
% \end{figure}

\begin{figure}
    \centering
    \includegraphics[width=1\linewidth]{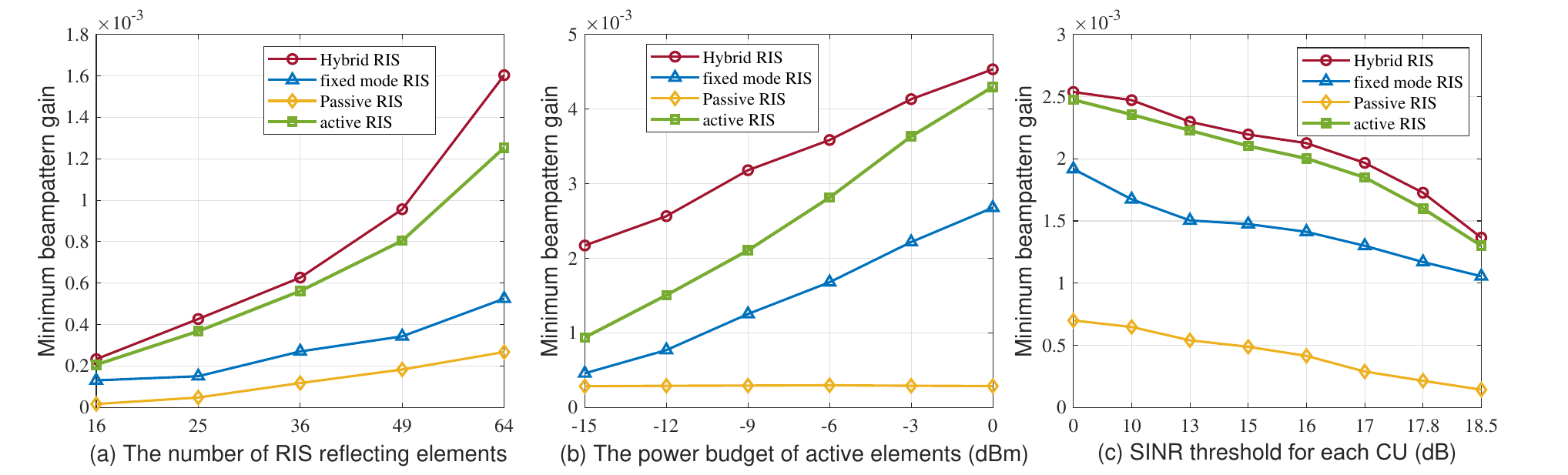}
    \caption{The achieved minimum sensing beampattern gain for target sensing: (a) versus the RIS element number $N$; (b) versus the RIS transmission power budget $P^{\rm{ris}}_{\rm{max}}$; (c) versus CUs' SINR threshold $\Gamma$.}
    \label{fig3}
    \vspace{-0.2cm}
\end{figure}

\begin{figure}[h]
	\begin{minipage}{0.49\linewidth}
		\vspace{3pt}
        %这个图片路径替换成你的图片路径即可使ç"?
		\centerline{\includegraphics[width=\textwidth]{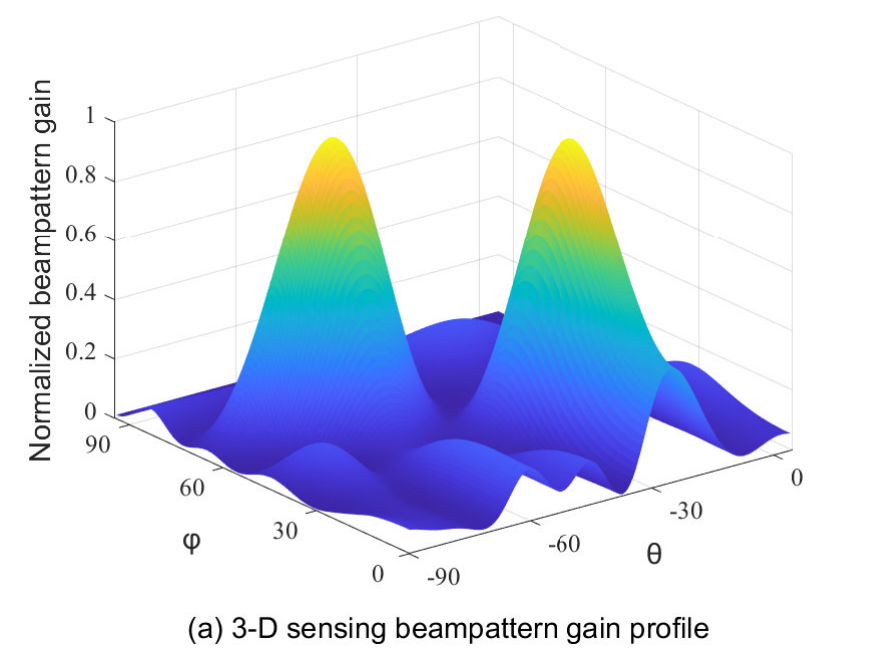}}
          % 加入对这列的图片说明
	\end{minipage}
	\begin{minipage}{0.5\linewidth}
		\vspace{3pt}
		\centerline{\includegraphics[width=\textwidth]{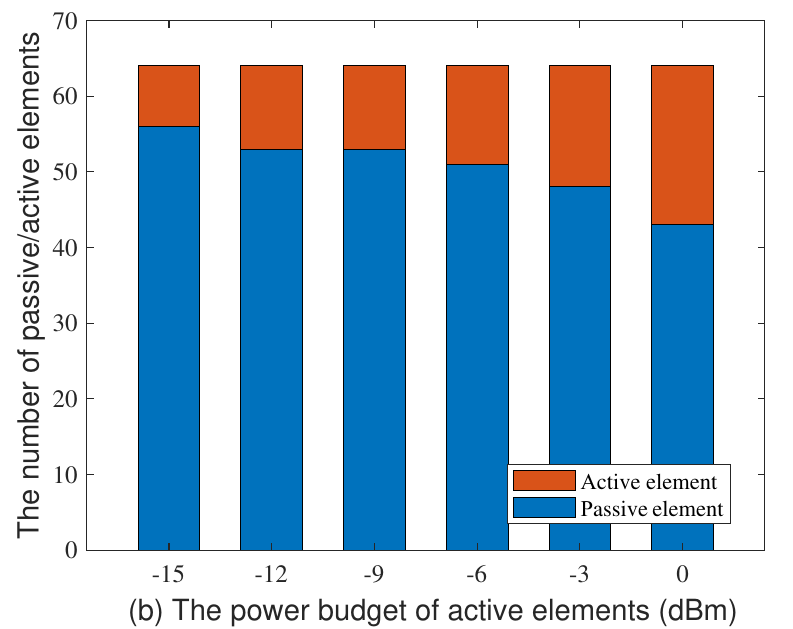}} 
	\end{minipage}
	\caption{Performance evaluation with $L=2$ targets and $K=2$. (a) The normalized 3-D sensing beampattern gain profile. (b) The number ratio of active-to-passive RIS elements.}
	\label{fig4}
    \vspace{-0.2cm}
\end{figure}

Fig.~\ref{fig3}(a) shows the minimum sensing beampattern gain versus the RIS element number $N$. As expected, the sensing beampatter gain increases with $N$. The proposed joint design scheme is observed to outperform the baseline schemes, and the performance benefit becomes significant in the case with a large RIS element number $N$. Fig.~\ref{fig3}(b) shows the minimum sensing beampattern gain versus the RIS transmission power budget $P_{\max}^{\text{ris}}$. As expected, the performance of all schemes improves as $P_{\max}^{\text{ris}}$ increases, and the proposed design scheme is observed to outperform the baseline schemes. Fig.~\ref{fig3}(c) shows the minimum sensing beampattern gain versus the SINR threshold $\Gamma$, and the proposed scheme achieves the best performance among all the schemes. The minimum sensing beampattern gains of all the schemes decrease with the increasing of $\Gamma$. This is because the CUs require a larger part of the transmission power with a larger value of $\Gamma$.

Fig.~\ref{fig4}(a) shows the 3-D sensing beampattern gain profile with $K=2$ and $L=2$. As expected, two peaks are observed in Fig.~\ref{fig4}(a), which clearly corresponds to the locations of the two targets to be sensed. This indicates that a significant portion of power is required to concentrate at the targets' positions. Fig.~\ref{fig4}(b) shows the number ratio of active-to-passive RIS elements under different setups of the RIS power budget $P_{\max}^{\text{ris}}$. It is observed that an increasing number of RIS elements choose the active mode as $P_{\max}^{\text{ris}}$ increases. It implies the importance of the RIS output power in determining the RIS active/passive mode selection.

\vspace{-0.3cm}
\section{Conclusion}
In this letter, we investigated a joint RIS mode selection and beamforming design for ISAC. Aiming to maximize the minimum sensing beampattern gain for targets, we jointly optimized the BS beamforming and the RIS active/passive mode selection, as well as the RIS reflection matrix. A low-complexity alternating optimization based solution was proposed to such joint RIS-assisted ISAC system design problem, where the SDR technique along with the rank-one beamforming reconstruction is employed at the BS, and the SDR and SCA techniques are employed for jointly optimizing RIS mode selection and reflection matrix. Numerical results showed the performance enhancement of the proposed design schemes when compared to the existing baseline schemes. %By carefully managing the active and passive mode of elements within RIS, this approach can be effectively deployed across various scenarios, making it suitable for a wide array of emerging applications.

\vspace{-0.3cm}
\bibliographystyle{IEEEtran}

% \bibliography{ref}

\end{document}